\documentclass{siamart190516}
\usepackage{graphicx, epsfig}
\usepackage{amsmath}
\usepackage{latexsym}
\usepackage{amstext}
\usepackage{epsfig}
\usepackage{latexsym}
\usepackage{amstext}
\usepackage{ulem}
\usepackage{amssymb}
\usepackage{array}
\usepackage{multirow}
\usepackage{tikz}
\usepackage{authblk}
\usepackage{cite}
\usepackage{amsmath}

\usepackage{amssymb}
\usepackage{array}
\usepackage{multirow}
\usepackage{tikz}
\newcommand{\mb}[1]{ \mbox{\boldmath$#1$} }

\newcommand{\ds}{\displaystyle}
\newcommand{\beq}{\begin{eqnarray}}
\newcommand{\eeq}{\end{eqnarray}}
\newcommand{\beqq}{\begin{eqnarray*}}
\newcommand{\eeqq}{\end{eqnarray*}}
\newcommand{\p}{\partial}

\newcommand{\x}{\mbox{\boldmath$x$}}

\newcommand{\y}{\mbox{\boldmath$y$}}

\newcommand{\n}{\mbox{\boldmath$n$}}

\newcommand{\J}{\mbox{\boldmath$J$}}

\def\ds#1{\displaystyle{#1}}

\begin{document}
\pagestyle{plain}
\begin{center}
{\large \textbf{{{Extreme diffusion with point-sink killing fields for fast calcium signaling at synapses}}}}\\[5mm]
S. Toste$^1$  D. Holcman $^{2}$\footnote{Group of applied mathematics, computational biology and predictive medicine, IBENS-PSL Ecole Normale Superieure, Paris, France. $^2$ Churchill College, CB30DS, DAMPT U. of Cambridge, Cambridge, UK.}
\end{center}

\begin{abstract}
We study here the escape time for the fastest diffusing particle from the boundary of an interval with point-sink killing sources. Killing represents a degradation that leads to the probabilistic removal of the moving Brownian particles. We compute asymptotically the mean time it takes for the fastest particle escaping alive and obtain the extreme statistic distribution. These computations relies on an explicit expression for the time dependent flux of the Fokker-Planck equation using the time dependent Green's function and Duhamel's formula. We obtain a general formula for several point-sink killing, showing how they directly  interact. The range of validity of the present formula for the mean extreme times of the fastest is evaluated with Brownian simulations. Finally, we discuss some applications to the early calcium signaling at neuronal synapses.
\end{abstract}

\begin{AMS}
{60G70, 35K05, 60J70, 92-10}
\end{AMS}

\section{Introduction}
For more than a century, the time scale of molecular activation has relied on the Smoluchowski's computation for the flux of a single particle reaching an absorbing sphere, a process modeled by the associated diffusion equation  \cite{chandrasekhar1943stochastic,Schuss2010,gardiner1985handbook,risken1996fokker}. This flux defines the reciprocal of the forward binding rate and also the time scale of cellular activation with a single molecular event. However, recently the time scale of activation for signaling event associated with calcium transients at neuronal synapses was found to be much faster than the one predicted by Smoluchowski's rate. This paradox about the fast time scale can be explained by the extreme statistical events \cite{schehr2014exact} for the arrival time of the fastest particles among many \cite{basnayake2019asymptotic,basnayake2019fast,basnayake2019fastest}. Briefly, there is no need of transporting a distribution of particles from one region to another to generate a response: the fastest arriving particles are sufficient to trigger the needed events after finding and binding to the key narrow targets. This event can for example open a channel that can trigger the release of the same species. This is well known in the case of calcium, known as calcium-induce-calcium-release \cite{dupont2016models}. The time of the fastest to arrive to a small target is in fact modulated by the initial copy number of identically distributed random particles. Recently, we hypothesize that this number sets the time to activation in most signaling molecular events, reproduction, gene expression and it is thus a fundamental achievement of life evolution at mostly all levels \cite{schuss2019redundancy,coombs2019first, martyushev2019minimal,metzler2014first, redner2019redundancy, rusakov2019extreme,sokolov2019extreme,tamm2019importance,ma2020strong}. Such large number guarantees that a rare event that would be impossible to trigger in a reasonable time scale will actually take place by the fastest particles in a reasonable time. This large number compensates for the unknown position of the small targets and the hidden geometry to be explored. The initial distribution of particles is often well separated from these target. This large number has been well calibrated for each applications, summarized as the redundancy principle \cite{schuss2019redundancy}.\\
The extreme statistics theory allows to compute the mean time of the fastest with respect to the parameters of the problem such as the diffusion coefficient for a diffusion process, the distance to the source and the initial number of particle
\cite{weiss1984perturbation,weiss1983order,yuste1996}. The computations have been extended to sub- and super- diffusion, but also when the initial distribution can extend close to the target window  \cite{grebenkov2020single,lawley2020distribution,lawley2020PRevE,toste2021asymptotics,linn2022extreme}.\\
In the present manuscript, we study the role of a killing source that can terminate the trajectory of a random particle before it can reach a target. The killing measure is the probability per unit time and unit length to terminate a trajectory. However, a moving particle can pass through a killing site many times without being terminated, in contrast to an absorbing boundary, where the trajectory is terminated with probability 1. Such a killing event can modify the escape time, due to the probability to be killed before escape \cite{holcman2005survival}. The probability of reaching small target and the conditional mean times are relevant to quantify the success of viral infection in cells \cite{lagache2009quantitative} or spermatozoa in the uterus \cite{reynaud2015so,meerson2015mortality,yang2016search}.\\
\begin{figure}[http!]
\begin{center}
\includegraphics[scale = 0.55]{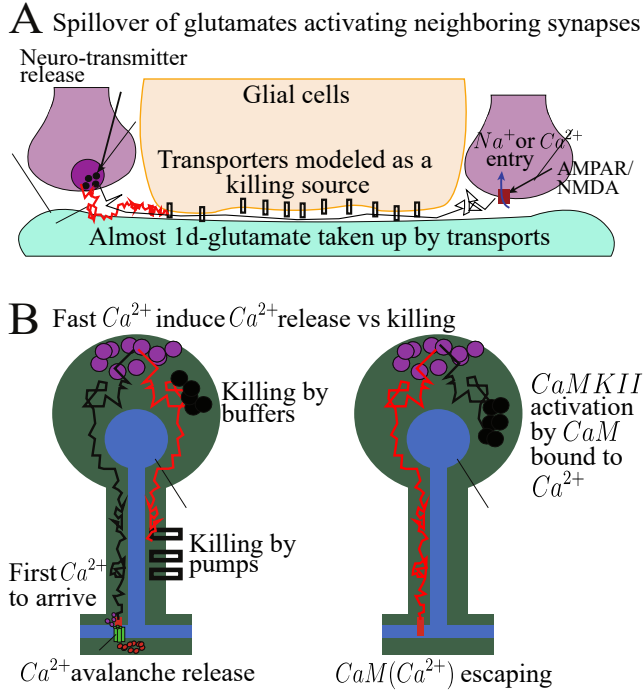}
\caption{\textbf{Escape versus killing in the regulation of molecular neuronal signaling.} \textbf{A.} Spillover of neurotransmitters after synaptic activation between glial cells (yellow), that contain transporters (modeled as killing term) to remove them from the extra-cellular space. Trajectory can be terminated (red) or arrive to receptor to activate the influx of ion in the neighboring synaptic terminal (purple).\textbf{B.} Left: Calcium dynamics in a dendritic spine: the fastest calcium ions can trigger a process called calcium-induce-calcium-release, if the ions are not stopped by a long-time binding buffer or extruded by pumps. Right: CaMKII kinase activation by calcium bound to calmodulin molecules. The probability and the mean time to activate the kinase CaMKII by the first calmodulin bound to calcium is one of the application explained here.}
\label{graph0}
\end{center}
\end{figure}
We are interesting here, in computing the mean time it takes for the fastest among many independent and equally distributed Brownian particles to reach a target when the killing measure is a sum of Dirac-delta functions located in an interval. To illustrate the present approach and the relevant of dimension reduction, we shall use two examples from neuroscience: the first one concerns the spillover of neurotransmitters such glutamate after synaptic activation. The neurotransmitters diffuse near glial cells that contains transporters (Fig.~\ref{graph0}A), the role of which is to remove these neurotransmitters from the extra-cellular space. This extrusion mechanism can be modelled as a one dimensional process with killing in an interval due the small space separation along the thin axone or dendrite. The second example concerns calcium dynamics in dendritic spines: the fastest calcium ions that enter following synaptic activation can trigger fast calcium release. However the fastest calcium ions should be interrupted by long-time binding buffers or extruded by pumps on their way to the base of the spine (Fig.~\ref{graph0}B). This interruption mechanism can be modeled by a killing term. We will also discuss below the case of calcium bound to calmodulin that can activate the CaMKII kinase. We propose to compute the probability and the mean time to activate a CaMKII \cite{nicoll2022}. This activation is relevant for the induction of long-term memory at a synaptic level. Here the relevant time is the first time that one CaM containing two calcium ions will arrive  at a CAMKII before it exits. This process is computed as the first bound to CaMKII, modeled by a killing measure.\\
In these examples, the role of the killing term is to terminate the particle trajectories at random times. The effect of the killing measure is accounted by an additive term in the Fokker-Planck equation, that describes the probability density function of the survival process before escape \cite{karlin1982diffusion, karlin1983class,berman1996distributions, holcman2005survival, mazzolo2022conditioning}.\\
The manuscript is organized as follows: in \cref{sec:theory}, we summarize the background: stochastic formulation and Fokker-Planck equation relevant to compute the mean first escape time under a killing field \cite{holcman2015stochastic}. In \cref{sec:main}, using a short-time asymptotic expansion of the diffusion equation with a single and multiples Dirac-delta killing measures, we derive a formula for the mean escape time for the first among many trajectories to escape before being killed in half-a-line. In  \cref{sec:simulation}, we discuss the asymptotic result with respect to the stochastic simulations. In \cref{sec:application}, we apply the present concept to model and determine the time of key calcium activation processes that can trigger long-term memory in dendritic spines.
\section{General background: killing measure versus survival probability}\label{sec:theory}
\subsection{Stochastic framework}
A stochastic process $\x(t)$ in the domain $\Omega$ satisfies the equation
\beq\label{eqstochastic}
d{\x}=\mb{b}(\x)\,dt+\sqrt{2}\mb{B}(\x)\,d{\mb{w}}(t)\hspace{0.5em},\mbox{for}\ \x\in\Omega,
\eeq
where $\mb{b}(\x)$ is a smooth drift vector, $\mb{B}(\x)$ is a diffusion tensor, and $\mb{w}(t)$ is a vector of independent standard Brownian motions. A killing measure $k(\x)$ is added in the domain $\Omega$ with boundary $\p\Omega=\p\Omega_a\cup\p\Omega_r$, where $\p\Omega_a$ is a small absorbing part and $\p\Omega_r$ is the reflecting boundary. The transition probability density function (pdf) of the process $\x(t)$ with killing and absorption is the pdf of trajectories that have neither been killed nor absorbed in $\p\Omega_a$ by time $t$,
\beq \label{pxTtau}
 p(\x,t\,|\,\y)\,d\x=\Pr\{\x(t)\in\x+d\x,\,\tau^k>t,\,\tau^e>t\,|\,\y\},\nonumber
\eeq
where $\tau^k$ is the time for the particle to be killed and $\tau^e$ is the time of absorbtion. This pdf is the solution of the Fokker-Planck equation (FPE) \cite{Schuss2010}
\beq \label{FPEp}
 \frac{\p p(\x,t\,|\,\y)}{\p t}={\cal L}_{\x} p(\x,t\,|\,\y)-k(\x)p(\x,t\,|\,\y)\hspace{0.5em}\ \mbox{for}
\ \x,\y\in \Omega,
 \eeq
where  ${\cal L}_{\x}$ is the forward operator
 \beq\label{FO}
{\cal
L}_{\x}p(\x,t\,|\,\y)=\sum_{i,j=1}^d\frac{\p^2\sigma^{i,j}(\x)p(\x,t\,|\,\y)}{\p
x^i\p x^j} -\sum_{i=1}^d\frac{\p b^i(\x)p(\x,t\,|\,\y)}{\p x^i},
 \eeq
and $\mb{\sigma}(\x)=\frac12\mb{B}(\x)\mb{B}^T(\x)$. The operator ${\cal
L}_{\x}$ can be written in the divergence form ${\cal
L}_{\x}p(\x,t\,|\,\y)=-\nabla\cdot\mb{J}(\x,t\,|\,\y)$, where the {components of the} flux density vector $\mb{J}(\x,t\,|\,\y)$ are
\beq
J^i(\x,t\,|\,\y)=-\sum_{j=1}^d\frac{\p\sigma^{i,j}(\x)p(\x,t\,|\,\y)}{\p
x^i}+b^i(\x)p(\x,t\,|\,\y),\hspace{0.5em}(i=1,2,\ldots,d).\nonumber 
\eeq
The initial and boundary conditions for the FPE (\ref{FPEp}) are
\beq
p(\x,0\,|\,\y)&=&\,\delta(\x-\y)\hspace{0.5em}\mbox{for}\
\x,\y\in \Omega\label{ICdD} \nonumber \\
p(\x,t\,|\,\y)&=&\,0\hspace{0.5em}\mbox{for}\  t>0,\
\x\in\p \Omega_a,\ \y\in \Omega\label{BCdD}\nonumber \\
\mb{J}(\x,t\,|\,\y)\cdot\n(\x)&=&\,0\hspace{0.5em}\mbox{for}\  t>0,\ \x\in\p
\Omega-\p \Omega_a,\ \y\in \Omega.\label{noflux} \nonumber
\eeq
The particular case where there is no drift vector, this is $b(x) = 0$, the FPE with the initial and boundary conditions written as above models the diffusive Brownian motion of particles that start at point $\y$. These particles are absorbed at point $x=0$ or degraded by the killing measure $k(\x)$.\\
The probability of trajectories that are killed before reaching $\p \Omega_a$ is given by \cite{Holcman2015},
\beq \label{Ttau}
\Pr\{\tau^k<\tau^e\,|\,\y\}=\int\limits_0^\infty\int\limits_{\Omega}
k(\x)p(\x,t\,|\,\y)\,d\x\,dt.\nonumber
\eeq
The absorption probability flux on $\p \Omega_{a}$ is
\beq \label{flux1}
J(t\,|\,\y)=\oint_{\partial
\Omega}\mb{J}(\x,t\,|\,\y)\cdot\n(\x)\,dS_{\x},
\eeq
and $\int_{0}^{\infty } J(t\,|\,\y)
\,dt$ is the probability of trajectories that have been absorbed at $\p \Omega_a$. Thus the probability to escape before being killed is
\beq\label{escapetau}
\Pr\{\tau^e<\tau^k\,|\,\y\}=\int_{0}^{\infty } J(t\,|\,\y)\,dt.
\eeq
The pdf of the killing time $\tau^k$ is the conditional probability of killing before time $t$ of trajectories that have not been absorbed in $\p \Omega_a$ by that time
\begin{align*}
\Pr\{\tau^k<t\,|\,\tau^e>\tau^k,\y\}=\frac{\Pr\{\tau^k<t,\tau^e>\tau^k\,|\,\y\}}
{\Pr\{\tau^e>\tau^k\,|\,\y\}}=\frac{\ds\int_{0}^{t}\int_{\Omega}k(\x)p(\x,s\,|\,\y)\,d\x\,ds}{\ds\int_0^\infty \int_{
\Omega}k(\x)p(\x,s\,|\,\y)\,d\x\,ds}.
\end{align*}
The probability distribution of the time to absorption at $\p \Omega_a$  is the conditional probability of absorption before time $t$ of trajectories that have not been killed by that time
\beq
\Pr\{\tau^e<t\,|\,\tau^k>\tau^e,\y\}=\frac{\ds\int_0^tJ(s\,|\,\y)\,ds}{1-\ds\int_0^\infty
\int_{\Omega}k(\x)p(\x,s\,|\,\y)\,d\x\,ds}.\nonumber
\eeq
Thus the narrow escape time (NET) is the conditional expectation of the absorption time of
trajectories that are not killed in $\Omega$, that is,
 \beq
\mathbb{E}[\tau^e\,|\,\tau^k>\tau^e,\,\y]=\,\int\limits_0^\infty\Pr\{\tau^e>t\,|\,\tau^k>\tau^e,\y\}\,dt=\,\frac{\ds \int_0^\infty s J(s\,|\,\y)\,ds}{1-\ds\int_0^\infty
\int_{ \Omega}k(\x)p(\x,s\,|\,\y)\,d\x\,ds}.\label{EtauT}\nonumber
 \eeq
The survival probability of trajectories that have not been terminated  by time $t$ is given by
\beq \label{survii}
S(t\,|\,\y)=\int\limits_{\Omega} p(\x,t\,|\,\y)\,d\x.
\eeq
For specific assumptions about the geometry of $\Omega$ and the distribution of absorbing windows, we refer to \cite{Holcman2015}.
\subsection{Extreme escape statistics with killing}
For $N_0$ independent identically distributed copies of the stochastic process (\ref{eqstochastic}), that can escape at time $t_1,...,t_{N_0}$, prior to get killed, we consider the escape time of the fastest one and we shall derive here a formula for the probability and mean escape time of the fastest Brownian motion. The extreme mean first passage time (EMFPT) $\bar\tau_{\footnotesize\mbox{EMFPT}}(n)$ \cite{lagache2009physical,Holcman2015} is the fastest time for a particle to escape through one of a narrow window located on the surface of the domain $\Omega$, that is
\beq
\tau_{\footnotesize\mbox{EMFPT}}(n)=\underset{n}{\min} \{ t_1,...,t_n \}. \nonumber
\eeq
All these times are conditioned to the fact that at least a large number of particles have to escape, so that $n \gg 1$ and $n \leq N_0$, where $n$ is the number of survival particles. The conditional mean first passage time (MFPT) ${\bar\tau}^j_n$ of the $j^{th}$ particle serves to compute $\tau_{EMFPT}(n)$ of the first particle that has reached the absorbing boundary $\p \Omega_a$. \\
The pdf of the escape time of the first particle prior to time $t$ with an initial density $p_0(x)$ is given by
\beq
P(t)=\Pr\{\tau^e_{EMFPT}(n)<t\,|\,\tau^e_{EMFPT}(n)<
\tau^k_{EMFPT}(n),p_0\}.\nonumber
\eeq
The conditional MFPT  $\bar\tau_{EMFPT}(n)$ is defined by
\beq
\bar\tau_{EMFPT}(n)  = \int\limits_0^{\infty} t \frac{dP(t)}{dt}\, dt =
\int\limits_0^{\infty}\left[P(\infty)-P(t)\right]\,dt\label{MFPTfirst}.
\eeq
Using Bayes' law, we obtain the decomposition
\beq
P(t) =\frac{\Pr\{\tau^e_{EMFPT}(n)<t,\tau^e_{EMFPT}(n)<\tau^k_{EMFPT}(n),p_0\}}
{\Pr\{\tau^e_{EMFPT}(n)<\tau^k_{EMFPT}(n),p_0\}}=\frac{N(t)}{P_{\infty}},\label{ND}
\eeq
where $P_{\infty}$ is the probability that the fastest one escape and the numerator $N(t)$ is defined as the conditional probability that the fastest one escapes alive before time $t$. Then, the extreme mean first passage time is conditioned to that at least one particle has to escape ($n\geq 1$).
\subsubsection{Probability that the fastest particle escapes}
The probability that the fastest particle escapes alive the domain is computed as follows
\beqq
P_{\infty}=\Pr\{\tau^e_{EMFPT}(n)<\tau^k_{EMFPT}(n),p_0\}=1-\Pr\{\tau^e_{EMFPT}(n)>
\tau^k_{EMFPT}(n),p_0\}.
\eeqq
Using that particles are independent, we get
\beqq
P_{\infty}=1-\prod_{j=1}^{n}\Pr\{\tau^e_j>\tau^k_j,p_0\},
\eeqq
which can be written as
\beq
P_{\infty}=1-\left(1-\Pr\{\tau^e<\tau^k,p_0 \}\right)^n.\nonumber
\eeq
According to relation (\ref{escapetau}), because the probability that a single particle escapes before being killed is given by $\Pr\{\tau^e<\tau^k,p_0\}=\int_{0}^{\infty } \int_{\y \in \Omega }J(t\,|\,\y)p_0(\y)d\y \,dt $ then,
\beq
P_{\infty}=1-\left(1-\int_{0}^{\infty } \int_{\y \in \Omega }J(t\,|\,\y)p_0(\y)d\y \,dt\right)^n.\nonumber
\eeq
For a Dirac-delta initial distribution at position $\y$, we get
\beq \label{probadirac}
P_{\infty}=1-\left(1-\int_{0}^{\infty } J(t\,|\,\y)dt\right)^n,
\eeq
where the flux $J$ is given by relation (\ref{flux1}).
Finally, the probability that $n-k$ particles are killed and only $k$ escape alive is given by the Binomial distribution
\beq
\Pr\{\tau^k<\tau^e,\tau^q>\tau^e,q= k+1,\ldots,n \}= \binom{n}{k} \left(\int_{0}^{\infty } J(t\,|\,\y)dt\right)^{k}\left(1-\int_{0}^{\infty } J(t\,|\,\y)dt\right)^{n-k}. \nonumber
\eeq
\subsection{Mean time for the fastest to escape without being killed}
The conditional probability that the fastest one escapes alive before time $t$ is given by
\beq
N(t)=\Pr\{\tau^e_{EMFPT}(n)<t,\tau^e_{EMFPT}(n)<\tau^k_{EMFPT}(n),p_0\},\nonumber
\eeq
that is,
\begin{eqnarray*}
\Pr\{\tau^e_{EMFPT}(n)<t,\tau^e_{EMFPT}(n)<\tau^k_{EMFPT}(n),p_0\} =\, 1-\Pr\{\tau^e_{EMFPT}(n)>t \hbox{ or }\tau^e_{EMFPT}(n)>\tau^k_{EMFPT}(n),p_0\}.    
\end{eqnarray*}
The event $\{\tau^e_{EMFPT}(n) >t \hbox{ or }\tau^e_{EMFPT}(n)>\tau^k_{EMFPT}(n)\}$ contains none of the $n$ particles that have escaped alive by time $t$. Because particles are independent, we obtain
\beqq
\Pr\{\tau^e_{EMFPT}(n) >t \hbox{ or
}\tau^e_{EMFPT}(n)>\tau^k_{EMFPT}(n),p_0\}
=\prod\limits_{j=1}^{n}
\left[1-\Pr\{\tau^e_j<t ,\tau^e_j<\tau^k_j,p_0\}
\right],
\eeqq
where $\tau^e_j$ (reps. $\tau^k_j$) is the first time that the $j^{th}$ particle is absorbed (resp. killed). Because the normal flux density at the boundary is the pdf of the exit point \cite{Schuss2010}, we get that for any of the particles
\beq \label{fluxx}
\Pr\{\tau^e_j<t ,\tau^e_j<\tau^k_j,p_0\} = \int\limits_0^t \oint\limits_{\partial
\Omega}\J(\x,t)\cdot\n(\x)\,dS_{\x}
=\int\limits_0^t J(s)\, ds,\nonumber
\eeq
where the flux $J(s)$ is defined in relation (\ref{flux1}). Therefore the numerator in equation \eqref{ND} is
\beq
N(t)=\,\Pr\{\tau^e_{EMFPT}(n)<t,\tau^e_{EMFPT}(n)<
\tau^k_{EMFPT}(n),p_0\}= 1-\left(1-\int\limits_0^t J(s) ds\right)^n. \nonumber
\eeq
To conclude, the conditional probability that the first particle, with an initial density $p_0(x)$ escapes alive at the absorbing boundary prior to time $t$ is given by
\beq \label{p(t)}
P(t) = \frac{N(t)}{P_{\infty}}=\frac{1-\left(1-\int_0^t J(s)\,
ds\right)^n}{1-\left(1-P_\infty^{(1)}\right)^n},
\eeq
where
\beq
P_\infty^{(1)}=\int_{0}^{\infty} J(s)\,ds, \nonumber
\eeq
and the conditional MFPT $\bar\tau_{EMFPT}(n)$ (see equation \eqref{MFPTfirst}) is
\beq
\bar\tau_{EMFPT}(n)=\int\limits_0^{\infty}\frac{\left(1-\int_{0}^{t} J(s)\,
ds\right)^n -\left(1-\int_{0}^{\infty} J(s)\,ds\right)^n}{1-\left(1-\int_{0}^{\infty} J(s)\,ds\right)^n}\, dt.\label{tau1}
\eeq
In \cite{lagache2009physical}, we previously derived a similar expression for the EMFPT, but we assumed that the survival probability decays exponentially. In the remaining part of the manuscript, we shall derive the full expression for the flux with a delta-Dirac killing source, without additional assumptions.\\
Similarly, we can compute the mean first killing time, given by the formula 
\beq
\bar\tau^k_{EMFPT}(n)  = \int\limits_0^{\infty} t \frac{dG(t)}{dt}\, dt =
\int\limits_0^{\infty}\left[G(\infty)-G(t)\right]\,dt\label{MFPTfirst_killing_event},
\eeq
where, 
\beq
G(t)=\Pr\{\tau^k_{EMFPT}(n)<t\,|\,\tau^k_{EMFKT}(n)<
\tau^e_{EMFPT}(n),p_0\}\nonumber
\eeq
is the probability of being killed before $t$, conditioned on the event that the particle is destroy or killed before escape. Proceeding as in formula \eqref{p(t)}, we obtain 
\beq
G(t) = \frac{1-\left(1-\int_{0}^{t} \int_{\Omega}k(x)p(x,s)\,dx\,
ds\right)^n}{\left(\int_{0}^{\infty}  \int_{\Omega}k(x)p(x,s)\,dx\,ds\right)^n},
\eeq
leading to the formula
\beq \label{form_killing}
\bar\tau^k_{EMFPT}(n)=\int\limits_0^{\infty}\frac{\left(1-\int_{0}^{t} \int_{\Omega}k(x)p(x,s)\,dx\,
ds\right)^n -\left(1-\int_{0}^{\infty}  \int_{\Omega}k(x)p(x,s)\,dx\,ds\right)^n}{\left(\int_{0}^{\infty}  \int_{\Omega}k(x)p(x,s)\,dx\,ds\right)^n}\, dt.
\eeq
\section{Extreme escape versus killing with a finite number of delta-Dirac isolated points}\label{sec:main}
\subsection{Survival probability with m-killing points}
We consider here $m$ isolated points in the half-a-line $\Omega = \mathbf{R}_+$ where diffusing particle can be degraded with a total weight $V=\sum_{i = 1}^{m}V_{i}$. The killing measure is given by
\beq
k(x)=\sum_{i=1}^{m}V_{i}\delta(x-x_i).\nonumber
\eeq
Brownian particles with diffusion coefficient $D$ can escape at the boundary $x= 0$.  To determine the formula for the fastest particle to escape alive, we solve the diffusion equation with the $m$ Dirac-killing terms by using the Green function \cite{holcman2005survival} in this domain. This method allows us to obtain an integral representation for the survival probability. The FPE is given by
\beq
\frac{\partial p(x,t\,|\,y)}{\partial t} &=& D
\frac{\partial^2p(x,t\,|\,y)}{\partial x^2}
-\sum_{i=1}^{m}V_{i}\delta(x-x_i)p(x,t\,|\,y) \label{edpdfm}\\
 p\left( x,0\,|\,y\right)  &=&\delta \left(
x-y\right) \nonumber
\\
 p\left( 0,t\,|\,y\right)  &=&0 . \nonumber
\eeq
This equation can be decomposed into:
\beq
\frac{\partial p(x,t\,|\,y)}{\partial t}-D\frac{\partial^{2}
p(x,t\,|\,y)}{\partial x^{2}}&=&F\label{edp1p}\\
p(x,0\,|\,y)& =& 0, \nonumber
\eeq
with $F =\displaystyle\sum_{i=1}^{m}V_{i}\delta(x-x_i)p(x,t\,|\,y),$ and
\beq
\frac{\partial p(x,t\,|\,y)}{\partial t}-D\frac{\partial^{2}
p(x,t\,|\,y)}{\partial x^{2}}&=&0 \label{edp2p}\\
p(x,0\,|\,y)& =& \delta(x-y) \nonumber
\\
p(0,t\,|\,y)& =& 0. \nonumber
\eeq
The fundamental solution of equation (\ref{edp2p}) is the heat kernel
\beq
G(x,t\,|\,y)=\frac{1}{2\sqrt{\pi Dt}}
\left(\exp\left\{-\displaystyle{\frac{(x-y)^{2}}{4Dt}}\right\}- \exp\left\{-\displaystyle{\frac{(x+y)^{2}}{4Dt}}\right\}\right), \nonumber
\eeq
while the solution of equation (\ref{edp1p}) is given by Duhamel's formula, in the form
\begin{eqnarray*}
 P(x,t\,|\,y)=  \int_{0}^{t}  \int_{\mathbb{R}} F(s,y)G(x,t-s\,|\,y)\,dy\,ds.
\end{eqnarray*}
Thus the general solution of equation (\ref{edpdfm}) is
\beq \label{edpsolmp}
&&p(x,t\,|\,y) = G(x,t)  \\
&-& \displaystyle \sum_{i=1}^m \int_0^t
\frac{V_ip(x_i,s\,|\,y)}{\sqrt{4\pi D(t-s)}}
 \left(\exp\left\{\displaystyle{\frac{-(x-x_i)^2}{4D(t-s)}}\right\} -\exp\left\{\displaystyle{\frac{-(x+x_i)^2}{4D(t-s)}}\right\}\right)ds. \nonumber
\eeq
The pdf $p(x,t\,|\,y)$ is known once the probability density functions $p(x_{1},t\,|\,y),\dots,p(x_{n},t\,|\,y)$ are determined. Setting $x=x_1$, $x=x_2$, ..., $x = x_m$ in equation (\ref{edpsolmp}) we obtain a system of integral equation in the single variable $t$ for the unknown functions
\beq
\phi_j(t)=p(x_j,t\,|\,y)  \hbox{ for } j = 1,\ldots,m.\nonumber
\eeq
We thus obtain
\beq
\phi_j(t) = G_j(t)- \sum_{i=1}^m \int_0^t \frac{ V_i \phi_i(s)}{\sqrt{4D\pi(t-s)}}\left(\exp\left\{-\displaystyle{\frac{(x_j-x_i)^2}{4D(t-s)}}\right\} -\exp\left\{-\displaystyle{\frac{(x_j+x_i)^2}{4D(t-s)}}\right\}\right) ds,\nonumber
\eeq
where $G_j(t) = G(x_j,t)$. The solution $p(x,t\,|\,y)$ will be determined once all the function $\phi_i(t)$ are known. To compute this, we use Laplace transform in time and we shall derive a system of linear equations
\beq \label{sysequation}
 \displaystyle \hat{\phi}_j(t) = \hat{G}_j(t)- \sum_{i=1}^m \frac{ V_i \hat{\phi}_i(s)}{\sqrt{4D\pi q}}\left(\exp\left\{-\displaystyle{\frac{|x_j-x_i|\sqrt{q}}{\sqrt{D}}}\right\} -\exp\left\{-\displaystyle{\frac{|x_j+x_i|\sqrt{q}}{\sqrt{D}}}\right\}\right).
 \eeq
Using the parameters $d_{ij} =\frac{|x_j-x_i|}{\sqrt{D}},\quad m_{ij} =\frac{|x_j+x_i|}{\sqrt{D}},\quad W_{i} = \frac{
V_{i}}{\sqrt{4D\pi}},$ we rewrite the system (\ref{sysequation}) in the matrix form
{\large
\beq
\mb{M}(x_{1},..,x_{m})\hat{\mb{\Phi}} =\hat{\mb{G}}, \nonumber
\eeq
}
where
\[ \mb{M}(x_{1},..,x_{m}) = \begin{pmatrix}
& 1+\displaystyle{W_{1} \frac{e^{-d_{11}\sqrt{q}}-e^{-m_{11}\sqrt{q}}}{\sqrt{q}}} \dots& W_{m}\displaystyle{\frac{e^{-d_{1m}\sqrt{q}}-e^{-m_{1m}\sqrt{q}}}{\sqrt{q}}} \\ &\hdots & \hdots\\ &\vdots & \vdots \\ &\dots& \dots\\ & W_{1}\displaystyle{\frac{e^{-d_{1m}\sqrt{q}}-e^{-m_{1m}\sqrt{q}}}{\sqrt{q}}} \dots & 1+\displaystyle{W_{m} \frac{e^{-d_{mm}\sqrt{q}}-e^{-m_{mm}\sqrt{q}}}{\sqrt{q}}}
\end{pmatrix} \]
and
\[ \hat{\mb{\Phi}} = \left(\begin{array}{c}{\hat \phi_1}\\ \vdots\\ \hat{\phi_{m}}
\end{array}\right),\, \hat{\mb{G}} = \left (
\begin{array}{c}
\hat{G_{1}} \\ \vdots\\ \hat{G_{m}}
\end{array}\right).
\]
We can write the matrix equation above as
\beq
\mb{M}(x_{1},..,x_{m}) = I_m+\frac{\mb{N}(x_{1},..,x_{m})}{\sqrt{q}},\nonumber
\eeq
where
\[ \mb{N}(x_{1},..,x_{m}) = \left[ W_{j}\left( e^{-d_{ij}\sqrt{q}}-e^{-m_{ij}\sqrt{q}}\right)\right]_{ij} \]
for $i,j = 1,\ldots,m$ and the coefficients of $\mb{N}(x_{1},..,x_{m})$ are algebraic functions of $d_{ij}$ and $m_{ij}$ depending on the Laplace variable $q$. The matrix $\mb{M}(x_{1},..,x_{m})$ is the sum of the identity with an $O\left(\frac{1}{\sqrt{q}}\right)$ perturbation, thus it is invertible and for $q$ large, we have the formal expansion
\beq \label{inversion}
\mb{M}^{-1}(x_{1},..,x_{m}) = \left( I_m + \frac{\mb{N}(x_{1},..,x_{m})}{\sqrt{q}} \right)^{-1} = \sum_{k=0}^{\infty} \left(-\frac{\mb{N}(x_{1},..,x_{m})}{\sqrt{q}} \right)^k \approx I_m -\frac{\mb{N}(x_{1},..,x_{m})}{\sqrt{q}}.\nonumber
\eeq
The solution can be written as  $\hat{\mb{\Phi}} =\mb{M}^{-1}(x_{1},..,x_{m})\hat{\mb{G}}$. We will use below the first order approximation of order $\frac{1}{\sqrt{q}}$ to estimate the leading order term of the mean extreme escape time. \\
We shall now compute the probability that the first particle escapes alive. Using relation (\ref{escapetau}), we have
\beq
\int_0^\infty J(s)ds=D\int_{0}^{\infty } \frac{\p p}{\p x}(x=0,t|\,y)\,dt. \nonumber
\eeq
Differentiating relation (\ref{edpsolmp}) and evaluating the Laplace's transform in $q=0$, we get
\beq
D\int_{0}^{\infty } \frac{\p p}{\p x}(x=0,t|\,y)\,dt= 1- \sum_{i=1}^m V_i\hat \phi_i(0). \nonumber
\eeq
Finally, using relation (\ref{probadirac}), we obtain for the escape probability
\beq
P_{\infty}= 1-\left(\sum_{i=1}^m V_i\hat \phi_i(0)\right)^n. \nonumber
\eeq
We shall now compute the EMFPT for the fastest Brownian particle. From formula (\ref{tau1}), we use a short-time expansion
\beq \label{numerator}
s(t)= \left(1-\int_{0}^{t} J(s)\,ds\right)^n-\left(1-\int_{0}^{\infty} J(s)\,ds\right)^n. \nonumber
\eeq
We then compute
\beq
\int_{0}^{t} J(s)\,ds &=& D\int_{0}^{t} \frac{\p p}{\p x}(x=0,s|\,y)\,ds \nonumber \\
&=& D \int_0^t \frac{\partial G}{\partial x}\left(x=0,s|\,y\right)ds -D \sum_{i=1}^m V_i \int_0^t \int_0^s \phi_i(u)\frac{\partial G}{\partial x}\left(x=0,s-u|\,x_i\right)\,du\,ds \nonumber \\
&=& \mathrm{erfc}\left(\frac{y}{\sqrt{4Dt}}\right) -D \sum_{1=1}^m V_i\int_0^t \phi_i(u) \mathrm{erfc} \left( \frac{x_i}{\sqrt{4D(t-u)}}\right)du.  \nonumber
\eeq
For $t$ small, the order of the integral
\beq
F_i(t) = D V_i\int_0^t \phi_i(u) \mathrm{erfc} \left( \frac{x_i}{\sqrt{4D(t-u)}}\right)du, \nonumber
\eeq
depends on the order of the functions $ \phi_i(u)$ and $\mathrm{erfc} \left( \frac{x_i}{\sqrt{4D(t-u)}}\right)$ that are continuous and differentiable functions in $[0,t]$ and $(0,t)$ respectively. Then, there exists a constant $c(t)\in [0,t]$ such that
\beq
F_i(t) = D V_i \phi_i(c_i(t)) \mathrm{erfc} \left( \frac{x_i}{\sqrt{4D(t-c_i(t))}}\right) t, \nonumber
\eeq
and, thus for $t$ small, $c_i(t)$ is small, and using the expansion for large argument of the $\mathrm{erfc}(x)$, we have the approximation,
\beq
F_i(t) = O\left(\exp \left \lbrace -\frac{x_i^2}{4D( t-c_i(t))} \right\rbrace \sqrt{(t-c_i(t))} t^{1+k}\right), \nonumber
\eeq
where $k$ is the order of $\phi_i(c_i(t))$. We have $\phi_i(0) = 0$ for $x_i \neq y$. When $x_i = y$, we have $\phi_i(0) = 1$ and
\beq
F_i(t) = O\left(\exp \left \lbrace -\frac{x_i^2}{4Dt} \right\rbrace t^{\frac{3}{2}+k} \right)>O\left(\exp \left \lbrace -\frac{x_i^2}{4Dt} \right\rbrace t^{\frac{1}{2}}\right). \nonumber
\eeq
Then, for $t$ small, the short-time asymptotic of $s(t)$ is dominated by the short-time asymptotic of
\beq
D \int_0^t \frac{\partial G}{\partial x}\left(x=0,s|\,y\right) = \mathrm{erfc}\left(\frac{y}{\sqrt{4Dt}}\right) . \nonumber
\eeq
Finally, we obtain from relation (\ref{tau1}),
\beq
\bar\tau_{EMFPT}(n)\sim \int\limits_0^{\infty} \frac{\left(1-\frac{\sqrt{4Dt} \exp \left \lbrace -\frac{y^2}{4Dt} \right\rbrace}{y \sqrt{\pi}}\right)^n -\left(\sum_{i=1}^m V_i\hat \phi_i(0)\right)^n}{1-\left(\sum_{i=1}^m V_i\hat \phi_i(0)\right)^n}\, dt.\nonumber
\eeq
Thus for $t$ small when $n$ large, we obtain
\beq
\bar\tau_{EMFPT}(n)&\sim& \int\limits_0^{\delta} \left[1-n \frac{\sqrt{4Dt} \exp \left \lbrace -\frac{y^2}{4Dt} \right\rbrace}{y \sqrt{\pi}\left(1-\left(\sum_{i=1}^m V_i\hat \phi_i(0)\right)^n\right)}\right]\, dt \nonumber \\
&\sim& \int\limits_0^{\infty} \exp \left\lbrace -n \frac{\sqrt{4Dt} \exp \left \lbrace -\frac{y^2}{4Dt} \right\rbrace}{y \sqrt{\pi}\left(1-\left(\sum_{i=1}^m V_i\hat \phi_i(0)\right) ^n\right)} \right\rbrace \, dt,
\eeq
and proceeding as in \cite{basnayake2019asymptotic}, we get
\beq \label{m_killing_points}
\bar\tau_{EMFPT}(n)&\sim& \frac{y^2}{4D \log \left( \frac{n}{\sqrt{\pi}\left(1-\left(\sum_{i=1}^m V_i\hat \phi_i(0)\right)^n\right)}\right)}.
\eeq
Formula \ref{m_killing_points} shows how the mean first escape time for the fastest depends on the various parameters. We shall now compute to leading order the term
\beq
T(V_1,...,V_n)=\sum_{i=1}^m V_i\hat \phi_i(0),\nonumber
\eeq
with respect with the physical parameters. Using the inverse matrix (\ref{inversion}), the first approximation gives
\beq
\hat \phi_i= \sum_{j} (I_m -\frac{\mb{N}(x_{1},..,x_{m})}{\sqrt{q}} )_{ij} \hat{G_{j}},\nonumber
\eeq
then,
\beq
\sum_{i=1}^m V_i\hat \phi_i(q)= \sum_{i,j}\left( V_i \hat{G_{i}}(q) -\frac{V_i V_{j}\alpha_{ij}(q)}{2\sqrt{D q}} \hat{G_{j}}(q)\right),\nonumber
\eeq
where $\alpha_{ij}(q)=e^{-d_{ij}\sqrt{q/D}}-e^{-m_{ij}\sqrt{q/D}}$. The Laplace transform of the Green's function is given by
\beq
\hat{G}(x_{i},q\,|\,\y)= \frac1{2\sqrt{Dq}}\left(\exp\left\{-|y
-x_i| \displaystyle{\sqrt{\frac{q}{D}}}\right\}-\exp\left\{-|y +x_i| \displaystyle{\sqrt{\frac{q}{D}}}\right\}
\right).\nonumber
\eeq
To conclude for $q=0$, we get
\beq \label{formulaT}
T(V_1,...,V_n)&=&\sum_{i=1}^m \frac{V_i}{2D}(|y-x_i| -|y+x_i| )-\sum_{i,j=1}^m \frac{V_j V_i}{2D^2}(|y-x_i| -|y+x_i| )(d_{ij}-m_{ij}).
\eeq
Formula (\ref{formulaT}) reveals the nonlinear dependency between the delta-Dirac located at position $x_i$ and the initial position $y$, the killing weights $V_i$ and the diffusion coefficient $D$. This term $T(V_1,...,V_n)$ is always less than 1. Consequently, for large $n$, it does not influence critically formula \eqref{m_killing_points} since it appears in the logarithmic term. We will exemplify this point more clearly in the next subsection where we only have one killing point.
\subsection{Survival probability with a single Dirac-delta killing measure}
We compute here the time-dependent survival probability (\ref{survii}) and the EMFPT for first among $n$ survival particles in the presence of a single Dirac-delta killing measure at position $x_1$ located on the half-line $x>0$. We recall that the FPE is given by
\beq \label{eq_id_1kp}
\frac{\partial p(x,t\,|\,y)}{\partial t} &=& D
\frac{\partial^2p(x,t\,|\,y)}{\partial x^2}
-V_{1}\delta(x-x_1)p(x,t\,|\,y) \label{edpdf1}\\
 p\left( x,0\,|\,y\right)  &=&\delta \left(
x-y\right) \nonumber
\\
 p\left( 0,t\,|\,y\right)  &=&0 . \nonumber
\eeq
The general solution of equation (\ref{edpdf1}) is the integral equation
\beq\label{edpsol2p}
&&p(x,t|y) = G(x,t\,|\,y) - \int_{0}^{t} \frac{V_{1}p(x_1,s\,|\,y)}
{2\sqrt{\pi
D(t-s)}}\left( \exp\left\{\displaystyle{\frac{-(x-x_1)^2}{4D(t-s)}}\right\} -\exp\left\{-\displaystyle{\frac{(x+x_1)^{2}}{4D(t-s)}}\right\} \right)ds.
\eeq
Setting $x=x_1$ in equation (\ref{edpsol2p}) reduces it to an integral equation in the single variable $t$ for the unknown function $\phi(t)=p(x_1,t\,|\,y)$. The solution $p(x,t\,|\,y)$ is completely determined once $\phi(t)$ is known. To compute this term, we use Laplace transform in time.  The integral equation (\ref{edpsol2p}) becomes
\beq
\hat\phi(q) &=& -V_{1}\frac{\hat\phi(q)}{2\sqrt{Dq}}\left(1-   \exp\left\{-|x_1| \displaystyle{\sqrt{\frac{2q}{D}}}\right\} \right)
+\hat{G}(x_{1},q\,|\,\y),\label{edpsol3}\nonumber
\eeq
where
 \beq
 \hat{G}(x_{1},q\,|\,y)= \frac1{2\sqrt{Dq}}\left(\exp\left\{-|y
-x_1| \displaystyle{\sqrt{\frac{q}{D}}}\right\}-\exp\left\{-|y +x_1| \displaystyle{\sqrt{\frac{q}{D}}}\right\}
\right).\nonumber
 \eeq
The solution is
\beq
\hat\phi(q) &=&\displaystyle{\frac{ \hat{G}(x_{1},q\,|\,y)}{ 1+
\displaystyle{\frac{ V_{1} }{2\sqrt{Dq}} } \left(1-   \exp\left\{-|x_1| {\sqrt{\frac{2q}{D}}}\right\} \right)  } }\nonumber =\frac{\left(\exp\left\{-|y
-x_1| {\sqrt{\frac{q}{D}}}\right\}-\exp\left\{-|y +x_1| {\sqrt{\frac{q}{D}}}\right\}\right)}{V_{1}\left(1-   \exp\left\{-x_1 {\sqrt{\frac{2q}{D}}}\right\} \right)+2\sqrt{Dq}}.\label{edpsolin}\nonumber
\eeq
We have
\beq \label{phizero}
\hat\phi(0) =\frac{|y +x_1| -|y -x_1|}{ V_{1} 2x_1+2D}.
\eeq
When $\hat\phi(q)$ is known, we obtain the general solution of (\ref{edpsol2p}) as
\beq\label{edpsolgeneral}
&&\hat p(x,q\,|\,y) = \hat{G}(x,q\,|\,y)
-V_{1}\frac{\hat\phi(q)}{2\sqrt{Dq}}\left(\exp\left\{-|x-x_1| \displaystyle{\sqrt{\frac{q}{D}}}\right\} -   \exp\left\{-|x+x_1| \displaystyle{\sqrt{\frac{q}{D}}}\right\} \right), \nonumber
\eeq
and thus,
\beq \label{expression}
\hat p(x,q\,|\,y) &=& -\frac{ V_{1}}{4Dq+V_1\sqrt{4Dq}\left(1-   \exp\left\{-2|x_1| {\sqrt{\frac{q}{D}}}\right\} \right)} \left(\exp\left\{-(|y -x_1| +|x-x_1|){\sqrt{\frac{q}{D}}}\right\} \right.\nonumber \\
&-& \left.\exp\left\{-(|y +x_1| +|x-x_1|){\sqrt{\frac{q}{D}}}\right\}
+\exp\left\{-(|y +x_1| +|x+x_1|){\sqrt{\frac{q}{D}}}\right\}\right.
 \\
&-&\left.\exp\left\{-(|y -x_1| +|x+x_1|){\sqrt{\frac{q}{D}}}\right\}\right)+\hat{G}(x,q\,|\,y).\label{edpsolgeneral1}\nonumber
\eeq
We rewrite expression \ref{expression} as a sum of the five terms, that we shall compute separately:
\beq
\hat p(x,q\,|\,y) = \hat p_1(x,q\,|\,y)+\hat p_2(x,q\,|\,y)+\hat p_3(x,q\,|\,y)+\hat p_4(x,q\,|\,y)+\hat{G}(x,q\,|\,y).
\eeq
The first term is defined by
\beq
\hat p_1(x,q\,|\,y) = -\frac{V_{1}}{4D}\frac{\exp\left\{-(|y -x_1| +|x-x_1|){\sqrt{\frac{q}{D}}}\right\}
}{q+ \frac{V_1}{2\sqrt{D}} \sqrt{q}}, \nonumber
\eeq
We apply the inverse Laplace for each solution using the generic expression for $\alpha>0$,
\begin{eqnarray*}
{\cal L}^{-1}\left(\frac{e^{-\alpha
\sqrt{q}}}{q+\sqrt{q}\displaystyle{\frac{V_1}{2\sqrt{D}}}}\right)=
\exp\left\{\displaystyle{\frac{\alpha
V_1}{2\sqrt{D}}}+\displaystyle{\frac{V^{2}_{1}}{4D}}t\right\}
\mbox{erfc}\left(\frac{\alpha}{2t^{1/2}}+\frac{V_{1}}{2\sqrt{D}}t^{1/2}\right).
\end{eqnarray*}
We obtain
\beq
p_1(x,t,|\,y) &=& -\frac{V_{1}}{4D} \exp\left\{\displaystyle{\frac{(|y -x_1| +|x-x_1|)
V_1}{2D}}+\displaystyle{\frac{V^{2}_{1}}{4D}}t\right\} \mbox{erfc}\left(\frac{(|y -x_1| +|x-x_1|)}{\sqrt{4Dt}}+\frac{V_{1}}{2\sqrt{D}}t^{1/2}\right).\nonumber
\eeq
For $t \ll 1$, we have the expansion
\beq
p_1(x,t,|\,y) \approx -\frac{V_{1}}{4D} \exp\left\{\displaystyle{\frac{(|y -x_1| +|x-x_1|)
V_1}{2D}}\right\}
\mbox{erfc}\left(\frac{(|y -x_1| +|x-x_1|)}{\sqrt{4Dt}}\right), \nonumber
\eeq
similarly for the other term in relation \ref{expression}:
\beq
p_2(x,t,|\,y) \approx \frac{V_{1}}{4D} \exp\left(\displaystyle{\frac{(|y +x_1| +|x-x_1|)
V_1}{2D}}\right)
\mbox{erfc}\left( \frac{(|y +x_1| +|x-x_1|)}{\sqrt{4Dt}}\right), \nonumber
\eeq
\beq
p_3(x,t,|\,y) \approx -\frac{V_{1}}{4D} \exp\left\{\displaystyle{\frac{(|y +x_1| +|x+x_1|)
V_1}{2D}}\right\}
\mbox{erfc}\left(\frac{(|y +x_1| +|x+x_1|)}{\sqrt{4Dt}}\right), \nonumber
\eeq
\beq
p_4(x,t,|\,y) \approx \frac{V_{1}}{4D} \exp\left\{\displaystyle{\frac{(|y -x_1| +|x+x_1|)
V_1}{2D}}\right\}
\mbox{erfc}\left(\frac{(|y -x_1| +|x+x_1|)}{\sqrt{4Dt}}\right).\nonumber
\eeq
We shall now compute the probability that the first particle escapes alive. Using relation (\ref{escapetau}), we have
\beq
\int_0^\infty J(t)dt=D\int_{0}^{\infty } \frac{\p p}{\p x}(x=0,t|\,y)\,dt. \nonumber
\eeq
Differentiating relation (\ref{edpsolgeneral}) and evaluating in $q=0$, we get
\beq
D\int_{0}^{\infty } \frac{\p p}{\p x}(x=0,t|\,y)\,dt=1-V_{1}\hat\phi(0). \nonumber
\eeq
Finally, using relation (\ref{probadirac}) and (\ref{phizero}), we get
\beq
P_{\infty}= 1-(V_{1}\hat\phi(0))^n=1-\left(V_1\frac{|y +x_1| -|y -x_1|}{ V_{1} 2|x_1|+2D}\right)^n. \nonumber
\eeq
We shall now compute the EMFPT for the fastest. Using formula (\ref{tau1}), we obtain that the short-time asymptotic for
\beq
s(t)= \left(1-\int_{0}^{t} J(s)\,ds\right)^n-\left(1-\int_{0}^{\infty} J(s)\,ds\right)^n. \nonumber
\eeq
Indeed, using the expansion of the complementary error function for large argument, we get from relation (\ref{tau1}) that
\beq
\bar\tau_{EMFPT}(n)\sim \int\limits_0^{\infty} \frac{\left(1-\frac{\sqrt{4Dt} \exp \left \lbrace -\frac{y^2}{4Dt} \right\rbrace}{y \sqrt{\pi}}\right)^n -\left(V_1 \hat{\phi}(0)\right)^n}{1-\left(V_1 \hat{\phi}(0)\right)^n}\, dt.\nonumber
\eeq
This integral can be estimated for $n \gg1$ as
\beq \label{approx_mfpt}
\bar\tau_{EMFPT}(n)&\sim& \int\limits_0^{\delta} \left[1-n \frac{\sqrt{4Dt} \exp \left \lbrace -\frac{y^2}{4Dt} \right\rbrace}{y \sqrt{\pi}\left(1-\left(V_1 \hat{\phi}(0)\right)^n\right)}\right]\, dt \nonumber \sim \int\limits_0^{\infty} \exp \left\lbrace -n \frac{\sqrt{4Dt} \exp \left \lbrace -\frac{y^2}{4Dt} \right\rbrace}{y \sqrt{\pi}\left(1-\left(V_1 \hat{\phi}(0)\right) ^n\right)} \right\rbrace \, dt, \nonumber
\eeq
and proceeding as in \cite{basnayake2019asymptotic}, we get
\beq \label{formula1}
\bar\tau_{EMFPT}(n)&\sim& \frac{y^2}{4D \log  \left( \frac{n}{\sqrt{\pi}\left(1-\left(V_1 \hat{\phi}(0)\right)^n\right)}\right)} \nonumber \\
&\sim& \frac{y^2}{4D \left[\log \displaystyle \left( \frac{n}{\sqrt{\pi}}\right) - \log \left(1-\left(V_1\frac{|y +x_1| -|y -x_1|}{ V_{1} 2|x_1|+2D}\right)^n\right) \right] }.
\eeq
Remarkably, since $X=\left(V_1\frac{|y +x_1| -|y -x_1|}{ V_{1} 2|x_1|+2D}\right) = \frac{1}{1+\frac{D}{V_1 x_1}} < 1$, when $n$ is large, using  $-\log(1-X^n)\approx X^n$, we obtain to leading order
\beq\label{emfpt_form}
\displaystyle \bar\tau_{EMFPT}(n)&\sim& \frac{y^2}{4D \left[ \log \left( \frac{n}{\sqrt{\pi}}\right)+X^n\right]}.
\eeq
Formula (\ref{emfpt_form}) reveals that the killing term decreases the mean time for the fastest particle to escape but still the leading order term is given by the logarithmic law.\\
We can also compute the escape time distribution of the fastest particle
\beq \label{escape_time_dist_1_delta_dim1}
&&Pr\left\lbrace \bar{\tau}^1 = t\right\rbrace
= -\frac{d}{dt} S(t) \sim  -\frac{d}{dt}\left[\exp \left\lbrace \frac{- n \sqrt{4 D t} e^{-\frac{y^2}{(4Dt)}}}{y \sqrt{\pi}\left(1-(V_1\hat{\phi}(0))^n\right)}\right \rbrace \right]  \\
&\sim&  \frac{ n\sqrt{4 D t} e^{-\frac{y^2}{(4Dt)}}}{y \sqrt{\pi} \left(1-(V_1\hat{\phi}(0))^n\right)}\exp \left\lbrace  \frac{- n \sqrt{4 D t}e^{-\frac{y^2}{(4Dt)}}}{y \sqrt{\pi}\left(1-(V_1\hat{\phi}(0))^n\right)}\right \rbrace  \left[ \frac{1}{2t} + \frac{y^2}{(4Dt^2)} \right].\nonumber
\eeq
Equivalently, we can have the formula for the mean first killing time given by \eqref{form_killing}, where 
$$\left(1-\int_{0}^{\infty}  \int_{\Omega}k(x)p(x,s)\,dx\,ds\right)^n = V_1 \hat{\phi}_1(0) \hbox{ and }\int_{0}^{t} \int_{\Omega}k(x)p(x,s)\,dx\,ds \approx \frac{V_1}{4D}\frac{\exp\left\lbrace-\frac{(x_1-y)^2}{4Dt}\right\rbrace(4Dt)^{\frac{3}{2}}}{\sqrt{\pi}(x_1-y)^2}.$$
Thus, we obtain
\beq
\bar\tau ^k_{EMFPT}(n) \sim \int\limits_0^{\infty} \frac{\left(1-\frac{V_1(4Dt)^{\frac{3}{2}} \exp \left \lbrace -\frac{(x_1-y)^2}{4Dt} \right\rbrace}{4D(y-x_1)^2 \sqrt{\pi}}\right)^n -\left(1-V_1 \hat{\phi}(0)\right)^n}{\left(V_1 \hat{\phi}(0)\right)^n}\, dt.
\eeq
Computing asymptotically the integral above, we obtain the formula for the extreme mean first killing time
\beq \label{killing_1point_formula}
\bar\tau ^k_{EMFPT}(n) \sim \frac{\left(1-\left(1-V_1\hat{\phi}(0)\right)^{n}\right)(y-x_1)^2}{4D\left(V_1\hat{\phi}(0)\right)^{n} \left[\log \displaystyle \left( \frac{n V_1 (y-x_1)}{4D\sqrt{\pi}\left(1-\left(1-V_1\hat{\phi}(0)\right)\right)^n} \right) \right]}.
\eeq
\section{Applications: numerical simulations and quantifying calcium signaling events in synapse}\label{sec:simulation}
In this section, we study the range of validity of the asymptotic formula derived above. We also show how the diffusion with killing can be used to quantify calcium dynamics in a sub-cellular compartment called the spine neck \cite{yuste2010dendritic}.
\subsection{Stochastic simulations of the fastest with a prescribed and floating large number n} \label{ss:ssimulation}
\begin{figure}[http!]
\begin{center}
\includegraphics[scale = 0.9]{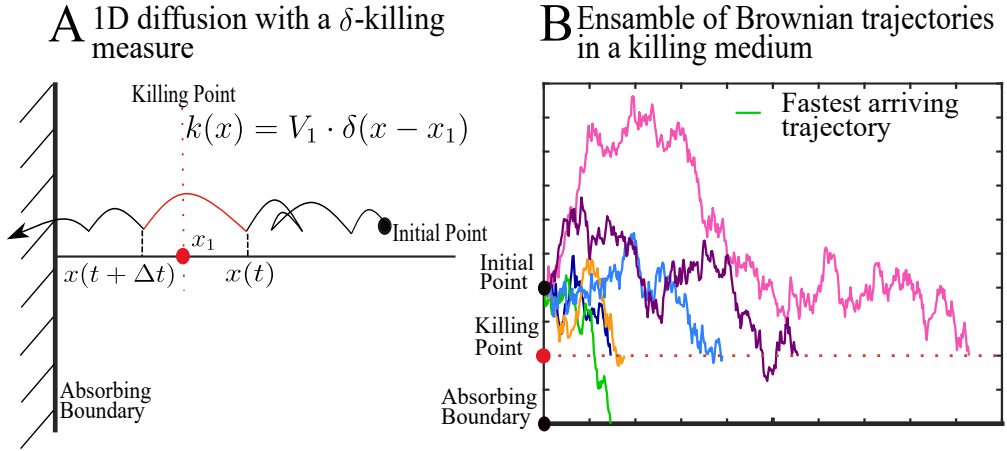}
\caption{\textbf{Escape versus killing for the fastest particles.} \textbf{A.} 1D Brownian motion passing through the  Dirac-delta killing field at point $x_1$. The particle is absorbed when reaching the boundary on the left. \textbf{B.} Five among six random walks are terminated while the extreme survival trajectory (green) reaches the boundary.}
\label{graph1}
\end{center}
\end{figure}

\begin{figure}[http!]
\begin{center}
\includegraphics[scale = 0.62]{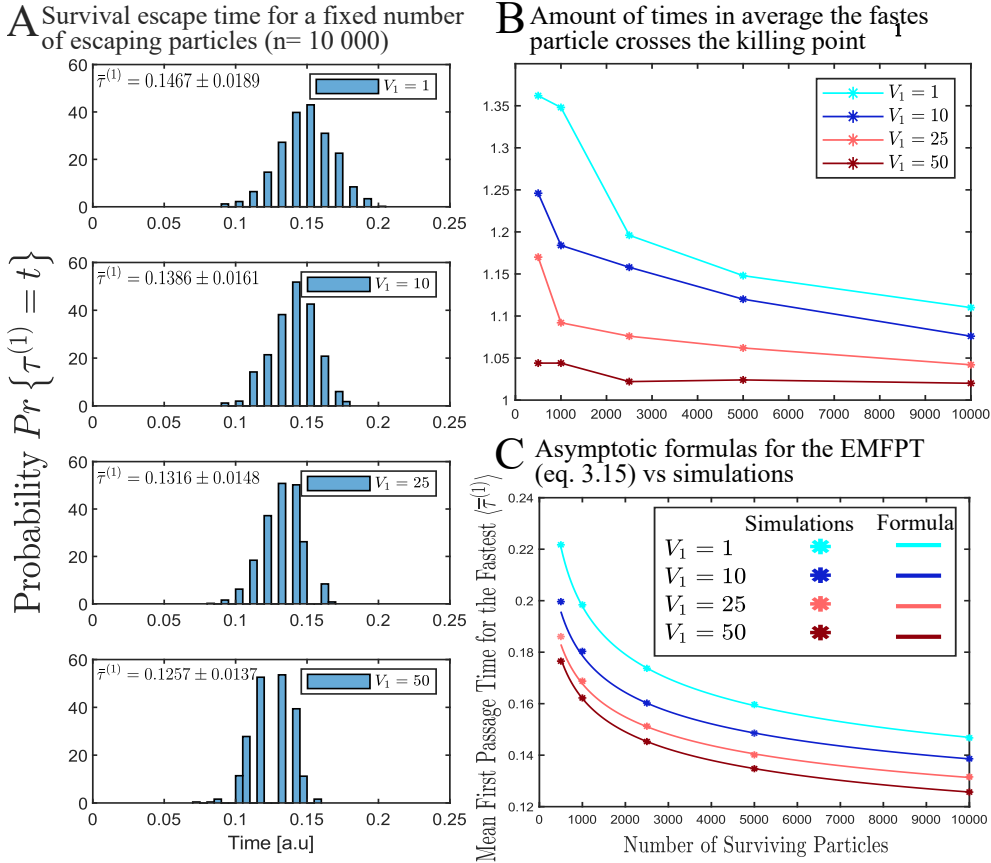}
\caption{\textbf{Influence of the killing rate on the mean escape time for the fastest particle.} \textbf{A.} Stochastic simulations for the escape time distribution of the fastest particle $\bar{\tau}^{1}$ for particles distributed with respect to $p_0 (x) = \delta(x-y)$  with $y = 2$ and a killing point in $x_1 = 1$ for $n=10 000$ with $1000$ runs. \textbf{B.} Decrease in the number of time the fastest particle crosses the killing point $x_1 = 1$ with the increasing of the killing weight for 1000 runs. \textbf{C.} EMFPT vs $n$ obtained from stochastic simulations (colored disks) and the asymptotic formulas (continuous lines) with $y = 2$, $x_1 = 1$ and 1000 runs.}
\label{graph2}
\end{center}
\end{figure}

\begin{figure}[http!]
\begin{center}
\includegraphics[scale = 0.56]{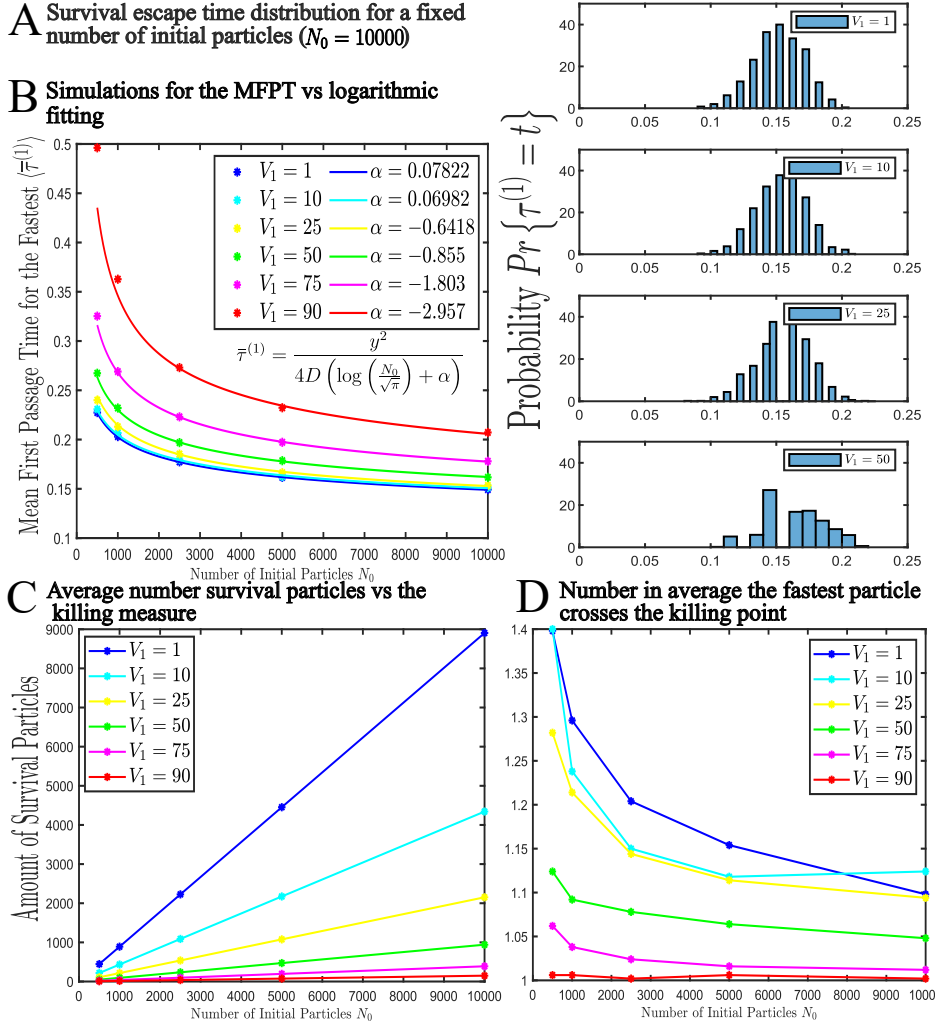}
\caption{\textbf{Influence of the killing rate on the escape time for a large number $N_0 \gg1$  of initial particles.} \textbf{A.} Stochastic simulations for the escape time $\bar{\tau}^{1}$ distribution of the fastest  particles, when the initial distribution is $p_0 (x) = \delta(x-y)$  with $y = 2$ and a killing measure at point $x_1 = 1$ for $N_0=10 000$ with $1000$ runs. \textbf{B.} EMFPT vs $N_0$ obtained from stochastic simulations (colored disks) and the asymptotic formulas (continuous lines) with $y = 2$, $x_1 = 1$ and 1000 runs. \textbf{C.} Influence of the killing weight $V_1$ in the number of survival particles. \textbf{D.} Decay of the number of time the fastest particle crosses the killing point $x_1 = 1$ when the killing weight increases (1000 runs).}
\label{graph3}
\end{center}
\end{figure}

We discuss here several applications of the EMFPT computations presented above. First, to test the range of accuracy of the asymptotic formulas, we run stochastic simulations for the first escape time with a killing Dirac-delta at point $x_1$ when all particles are initially distributed at position $y$ modeled as $p_0(x)=\delta(x-y)$ for different number $n$ of particles and killing weight $V_1$.
The stochastic simulation follows Euler's scheme (Fig.~\ref{graph1}A): for a particle crossing the point $x_1$ in any sense during the time step $\Delta t$, that is $x(t)\leq x_1 \leq x(t+\Delta t)$ or the other side, we  have
\beq
x(t+ \Delta t) =\left\lbrace
\begin{matrix}
x(t) + \sqrt{2D} \Delta w(t) & \text{w.p $1-V_1I_{\{x(t) \leq x_1 \leq x(t+\Delta t)\} \text{ or } \{x(t+\Delta t) \leq x_1 \leq x(t)\}} \Delta t$} \\ &  \\
\hbox{TERMINATED}, & \text{w.p $V_1I_{\{x(t) \leq x_1 \leq x(t+\Delta t)\} \text{ or } \{x(t+\Delta t) \leq x_1 \leq x(t)\}} \Delta t$}
\end{matrix}\right. \nonumber
\eeq
Live particles can be destroyed at Poissonian rate $V_1$ with probability $V_1\Delta t$, when passing over the point $x_1$ \cite{erban2019stochastic,oshanin2019}. We are interested in the statistical properties of the fastest particle reaching the absorbing boundary prior to be killed (Fig.~\ref{graph1}B).
Outside the crossing point $x_1$, the Euler's scheme is the classical Brownian jump at scale $\Delta t$. We started the simulation at point $y =2$ with diffusion coefficient $D=1$ with the killing point at $x_1 = 1$, with a time step $\Delta t =0.01$.Note that we do not fix the initial number of particles $N_0$, but we run simulations until we reach a given amount $n$ of survival particles with $n = $ [500 1000 2500 5000 10000].
As shown in Fig.~\ref{graph2}A, the simulated mean escape time decays with the killing weight $V_1$ in agreement with formula (\ref{escape_time_dist_1_delta_dim1}). Interestingly, the fastest particles crosses the killing point only a few times and this number decreases when the killing weight increases (Fig.~\ref{graph2}B). After the fastest particles has crossed the killing zone, it does not cross it again. Finally, as the number of particles $n$ increases, the fastest particle moves directly toward the absorbing point to exit. The EMFPT decreases with the number of survival particles as illustrated in Fig.~\ref{graph2}C. In summary, the asymptotic formula (\ref{emfpt_form}) is robust over a large range of n and killing rate $V_1$, as confirmed by the agreement with the stochastic simulations.\\
We decided to further explore the consequence of fixing the initial number of particles $N_0 = $ [500 1000 2500 5000 10000], which does not necessarily correspond to the number of survival particles that will escape. In practice, much less particles will escape, thus reducing the total number used in the extreme statistics.  To illustrate this difference, we plotted the mean escape time versus the killing term (Fig.~\ref{graph3}A), and the EMFPT versus the killing probability (Fig.~\ref{graph3}B). The curves differs from the result shown in Fig.~\ref{graph2}, due to the decreasing in the number of survival particles. Such difference can be accounted for by adding a correction term $\alpha$ in the asymptotic formula for the EMFPT, as shown in Fig.~\ref{graph3}B. When the killing weight $V_1$ increases, the number of escaping particles $n$ decreases, as shown in Fig.~\ref{graph3}C. In that regime, the fastest particles also avoid crossing the killing point multiple times (Fig.~\ref{graph3}D).
\subsection{Time scale of fast calcium signaling at synapse}\label{sec:application}
Calcium dynamics at synapses is a fundamental step to transform neuronal spike coding, propagated across neurons into long-term molecular changes at a subcellular level, called synaptic plasticity, at the bases of learning and memory \cite{nicoll2022}. Interestingly, following a transient in the spine head (Fig.~\ref{graph4}), fast calcium increase in dendrite is much faster than predicted by the classical transport resulting from the theory of diffusion \cite{basnayake2019asymptotic}. This observation was interpreted as a consequence of the arrival of the fastest calcium ions that trigger calcium by a mechanism called calcium-induced-calcium-release through a class of receptor called Ryanodine receptor (RyR) located at the base of spine (Fig.~\ref{graph4}).
While the mean time of CICR was previously computed as the arrival of first two calcium ions to a RyR, this computed neglected the influence of calcium buffers that can capture calcium ion on their way for a long time, thus preventing a fast CICR.  The main calcium buffers in the cytoplasm includes Trophin C, Calmodulin, Calcineurin and Myosin. If the concentration of buffer is high, the calcium trajectory that will arrive to a target will be significantly reduced. Calcium buffers could thus prevent the fast activation of CICR or even a second messenger pathway such as IP3 receptors, located at the base of a spine \cite{sneyd1995mechanisms,dupont2017recent,smith2019calcium,holcman2005modeling}.
\subsubsection{Effect of calcium buffers modeled as a killing point source on Calcium-Induce-Calcium-Release}
\begin{figure}[http!]
\begin{center}
\includegraphics[scale = 0.45]{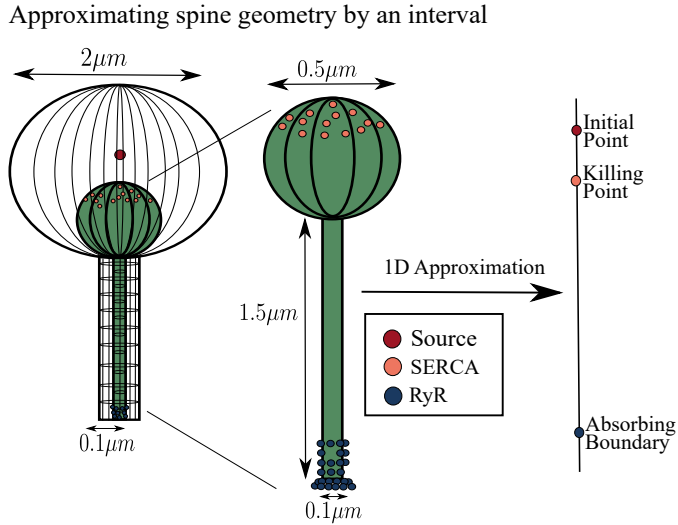}
\caption{\textbf{Schematic representation of a dendritic spine doted with a spine apparatus and its simplification in a 1D domain.} The spine with a spine apparatus is simplified as a 1D interval with killing point $x_1 = 2 \mu m$, initial point at $y =2.5 \mu m$, absorbing point $x=0 \mu m$.}
\label{graph4}
\end{center}
\end{figure}

We propose now to model calcium dynamics in spine head as a diffusion in narrow cylinder, approximated as a segment. Indeed, due to the small size of the narrow cylinder and head of the dendritic spine, we could approximate the motion of calcium particles inside the narrow cylinder by a one dimensional Brownian motion in an interval. The fast binding to a buffer molecule will be account for by killing term in the diffusion equation, and since unbinding is often much longer that the binding time (hundreds vs few milliseconds), we can neglect here the unbinding time. The cases of uniform killing measures occurring on a interval is discussed in appendix \cref{sec:appendix}. Some formula could be easily extended to the case of a partially absorbing target \cite{grebenkov2020single}. The effect of calcium removal by SERCA pumps can also be represented by a single or many killing points inside the interval $[0,L]$. The process of CICR induced by the binding of calcium ion to RyR is modeled as an absorbing boundary, where escape occurs.\\
We start the model, after there are a total of $n$ $Ca^{2+}$ ions that have entered the dendritic spine through the receptors (dark red point) located in the spine head (Fig.~\ref{graph4}). The time of CICR is computed after the arrival of two fastest $Ca^{2+}$ ions at the RyR (blue dots) at the bottom of the spine (absorbing boundary condition). After the RyR is activated, an avalanche through a CIRC from SA is generated. This leads to an amplification of the calcium signal.\\
The CICR process can be computed from the escape time distribution of the second fastest particle arriving to the absorbing end point of the interval, that model the spine neck. The pdf $Pr\left\lbrace \bar{\tau}^1 = s\right\rbrace$ for the time the first ion arriving to the boundary allows to compute the pdf for second one to arrive by conditioning on the arrival of the first one at time s, while there are still $n-1$ ions in the interval. Thus we obtain the relation:
\beq
Pr\left\lbrace \bar{\tau}^2 = t\right\rbrace
&=& \int_0^t Pr\left\lbrace \bar{\tau}^2 = t| \bar{\tau}^1 = s\right\rbrace S^{n-1}(s)  Pr\left\lbrace \bar{\tau}^1 = s\right\rbrace ds \\ &\sim&  \int_0^t Pr\left\lbrace \bar{\tau}^2 = t| \bar{\tau}^1 = s \right\rbrace Pr\left\lbrace \bar{\tau}^1 = s\right\rbrace ds,\nonumber
\eeq
where we consider that the remaining $n-1$ particles are still alive close to the initial position when the killing weight $V_1$ is not too large, thus we use the approximation \cite{basnayake2019asymptotic}
\beq
S^{n-1}(t) =  \left(\int_0^a Pr\left\lbrace x_2(t) = x_2\right\rbrace dx_2\right)^{n-1} \approx 1.\nonumber
\eeq
We approximate the motion inside the narrow cylinder by a one dimensional Brownian motion in an interval $[0,L]$, with $y<L$, where $y$ is the initial position of the source, as shown in  Fig.~\ref{graph4}A. In practice $y=L$. The buffer or SERCA pumps are represented by a single killing point.\\
Using the approximation summarized by equation (\ref{formula1}), and that $Pr\left\lbrace \bar{\tau}^2 = t| \bar{\tau}^1 = \textcolor{red}{s} \right\rbrace \approx Pr\left\lbrace \bar{\tau}^1 = t-s \right  \rbrace$,
the extreme escape time for the two fastest particles \cite{basnayake2019asymptotic} is computed directly, leading to
\beq \label{Cabufferarrival}
\bar\tau_{EMFPT}^2(n)\sim 2 \bar\tau_{EMFPT}(n) \sim \frac{2L^2}{4D \left[\log \displaystyle \left( \frac{n}{\sqrt{\pi}}\right) - \log \left(1-\left(V_1\frac{x_1}{ V_{1} |x_1|+D}\right)^n\right) \right] }.
\eeq
To conclude, relation \ref{Cabufferarrival} shows that the consequence of the killing buffer is to decrease the binding time and the probability $1-\exp (-\frac{nA}{V_1})$, where $A\approx \frac{D}{|x_1|}$. Interestingly, the formula shows a modulation depending on the position of the killing source. For several killing- delta-Dirac, the extreme mean first passage is given by formula (\ref{m_killing_points}). When buffer molecules are uniformly distributed, formula (\ref{formula4}) should be used instead.
\subsubsection{Probability and time to induce long-term change at a molecular level}
The second example we shall discuss consists in the molecular induction of plastic changes at a molecular level following high calcium concentration level entering into the neuronal synapse. The first step of the signaling consists in calcium ions binding to calmodulin and then the complex calcium-calmodulin needs to bind to a kinase third partner CaMKII \cite{nicoll2022}. We propose to estimate the probability to activate a given number $N_{KII}$ of CaMKII kinases inside a spine and how long does it takes for such activation. \\
We first consider that calcium bind quickly to calmodulin at the time scale given by the first ions to arrival to the molecule sites, of the order of less than 1 millisecond \cite{holcman2004calcium}.  The unbinding time is too long (hundreds compared to few milliseconds). The binding of CaM containing a calcium to the kinase can be achieved by the four components: $CaMCa_1$, $CaMCa_2$, $CaMCa_3$ and $CaMCa_4$. This can be summarized by the following chemical rate equations:
\beq
CAM+ Ca^{2+}     &\leftrightharpoons  CAMCa\\
CAMCa+ Ca^{2+}    &\leftrightharpoons  CAMCa_2\\
CAMCa_2+ Ca^{2+}  &\leftrightharpoons  CAMCa_3\\
CAMCa_3+ Ca^{2+}  &\leftrightharpoons  CAMCa_4.
\eeq
 We consider the approximation that the number of molecules in each category is given by $N_i = p_i n$, where $p_i=p^i$ with $i=1,...,4$, where $p<1$. Thus the number of bound CaM to calcium decays exponentially with the initial number of calcium ions. The complex $CaMCa_i$ can dissociate with a rate $\kappa$ which is much shorter than the binding rate. \\
We apply now the result developed in the previous section to $CaMCa^{2+}_i$ $i=1,...,4,$ that can diffuse and thus escape the spine at the absorbing boundary. In that case, using relation (\ref{phizero}), the probability that there are $N_{CaMKII}$ molecules of CaMKII bound by the population $CaMCa_i$ is given by the killing probability
\beq
P_i= \left( \frac{x_1 V_{1} }{ V_{1} x_1+D} \right)^{N_i}.
\eeq
Here we considered that the $CaMCa_i$ are located at position $x_1$ and $V_1$ represent the binding rate. When there are more CaMKII than CaM bound to calcium, then $V_1\approx k_1 N_{CaMKII}$, where $k_1$ is the forward binding rate. In general, the mean number of bound CaMKII can be computed using a binomial law associated to $P_i$. Thus
\beq
\langle CaMKII-CaMCa_i\rangle= P_i N_{CaMKII}
\eeq
and the variance is $P_i(1-P_i) N_{CaMKII}$. Finally, the total number of bound CaMKII is obtained by summing over $i=1,...,4$ as follows
\beq
\langle CaMKII_{\hbox{bound}}\rangle=\sum_i \langle CaMKII-CaMCa_i\rangle= \sum_i N_{ CaMKII} \left( \frac{ V_{1}  x_1}{ V_{1} x_1+D} \right)^{n p^i}.
\eeq
The time of activation of the $CaMKII$ molecules by the  population of $CaMCa_2$, which is the one that can lead to phosphorylation \cite{nicoll2022}, keeping the kinase $CaMKII$ active, is given in our model by the time for the first killing to occur, as it represents the binding of $CaMCa_2$ to $CaMKII$. This time can be computed from formula \eqref{killing_1point_formula} leading to
\beq
\bar\tau ^k_{EMFPT}(n) \sim \frac{\left(1-\left(\frac{\frac{D}{x_1 V_1}}{1+\frac{D}{x_1 V_1}}\right)^{p^2n}\right)(y-x_1)^2}{4D\left(\frac{1}{1+\frac{D}{x_1 V_1}}\right)^{p^2n} \left[\log \displaystyle \left( \frac{p^2 n V_1 (y-x_1)}{4D\sqrt{\pi}\left(1-\left(\frac{\frac{D}{x_1 V_1}}{1+\frac{D}{x_1 V_1}}\right)^{p^2n}\right)} \right) \right]},
\eeq
where $D=100 \mu m^2/s$, $n=50000$, $y=1 \mu m$, $x_1=0.1 \mu m$, $p=0.2$ \cite{holcman2004calcium}, $V_1$ is not known but it could be found from experiments. For instance if $k_1 = 50 \mu m/s$, we can find the mean time for activate the $CaMKII$ from replacing all this values in the formula, and thus we obtain $\bar\tau ^k_{EMFPT} = 0.0079s$, meaning that the activation of this molecules is in the order of a few milliseconds.
\section{Conclusions and perspective}\label{sec:conclusion}
We reported here various escape asymptotic laws for the fastest particles to reach the boundary of an interval when there are multiple delta-Dirac killing sources. We obtain asymptotic formula for the large number of particle limit. The formulas revealed the mixed role of dynamics and killing that influences the fastest particle to escape.\\
We used this framework to estimate how buffer can influence calcium dynamics at synapses in the process of calcium induce calcium release and the time of $CaMKII$ activation. In general, the present approach can be used to derive the time scale of biochemical processes, where signaling occurs through the fastest particles. This framework can also account for the time to activate an ensemble of chemical processes \cite{lu2006statistics} or the time for a chemical message to be delivered when it is carried by few particles among many \cite{sokolov2019extreme,basnayake2019fastestphys,coombs2019first}. Finding a target is key to activate sub-cellular process \cite{schuss2019redundancy}. However, during this event, the diffusing messenger can bind to molecules that can trap or destroy them, thus affecting the path of the fastest particles to their final target. These binding molecules can diminish the arrival probability, but interestingly, they reduce the time of arrival, as shown by formulas \ref{formula1}, \ref{formula2}, \ref{formula3}, \ref{formula4} and \ref{uniformkilling}: indeed, the fastest particles should avoid staying in the domain where they can terminated, easier with point wise or uniform killing distribution. These formula further reveal that the distribution of killing sources influences on the fastest escape time.\\
There are other examples where the present theory could be relevant: in the cell nucleus \cite{malherbe2010search}, transcription factors (TFs) are switching between different states before escaping to a small target site: the $TFs$ are moving as a Brownian particles and can bind to various ligands to change state (acethylation or sumolysation) \cite{alberts2013essential}. The TFs can be degraded, preventing the fastest to reach the target, while gene activation can only occur in one of the appropriate state. This example shows that the number of TFs can accelerate the production of mARN, but the escape time could be limited by killing processes. Finally, it would be interesting to extend the present study in higher dimensions where the fastest can avoid entering the killing region.
\section{Appendix}\label{sec:appendix}
We presented in this appendix the computations for the mean first escape time when the killing measure is uniform and located in an interval that may or may not contain the initial point.
\subsection{Escape for the fastest with a uniform killing in half-a-line}
We now consider the escape time for the fastest particle when the killing measure $k\left( x,t\right) =V_{0}$ is constant over the half-a-line $x\geq 0$. The diffusion
coefficient is $D$ and the survival FPE for each individual particle is
\beq \label{first}
\frac{\partial p(x,t\,|\,y)}{\partial t}&=& D\frac{\partial^{2} p(x,t\,|\,y)}
{\partial x^{2}} -V_{0}p(x,t\,|\,y),\quad \mbox{for $x\in\mathbb{R}_+,\ t>0$} \nonumber\\
p(x,0\,|\,y)& =& \delta(y-x) \nonumber \\
p(0,t\,|\,y)&=&0.\nonumber
\eeq
The solution of this equation is given by
\beq
p(x,t\,|\,y)=\exp\left\{-V_{0}t\right\}\frac{1}{2\sqrt{\pi Dt}}\left(
\exp\left\{-\frac{(x-y)^{2}}{4Dt}\right\}-\exp\left\{-\frac{(x+y)^{2}}{4Dt}\right\}\right)\label{k=V0}\nonumber
\eeq
and the flux is
\beq
J(t\,|\,y)=D \frac{\p p}{\p x}(x=0,t\,|\,y)=\exp\left\{-V_{0}t\right\}\frac{y}{t\sqrt{4 \pi Dt}}\left(
\exp\left\{-\frac{y^{2}}{4Dt}\right\}\right).\nonumber
\eeq
Thus using the inverse Laplace transform
\beq
\int_0^{\infty} \frac{1}{\sqrt{\pi}t^{3/2}}e^{-at-b/t} dt= \frac{1}{2\sqrt{b}}\exp \left \lbrace -2\sqrt{ab} \right \rbrace, \nonumber
\eeq
we find the expression for the probability to escapes alive for one particle
\beq
\int_0^{\infty} J(t\,|\,y)dt= \exp \left \lbrace -y\sqrt{\frac{V_{0}}{D}}\right \rbrace.\nonumber
\eeq
Thus, the probability that the first one escape alive in an ensemble of $n$ is
\beq
P_{\infty}=1-\left(1-\int_{0}^{\infty } J(t\,|\,y)dt\right)^n = 1-\left(1-\exp\left \lbrace -y\sqrt{\frac{V_{0}}{D}}\right \rbrace \right)^n.\nonumber
\eeq
Similarly, we obtain the expression for the total flux for a single particle
\beq
&&\int_0^{t} J(s\,|\,y)ds = \int_0^{t} \frac{y \exp \left \lbrace -V_0 s \right \rbrace \exp  \left \lbrace -\frac{y^2}{4Ds}\right \rbrace }{\sqrt{4D\pi s}s}ds \nonumber \\
&=& \frac{1}{2} \left(\exp\left \lbrace -y\sqrt{\frac{V_{0}}{D}}\right \rbrace\mathrm{erfc}\left( \frac{y}{\sqrt{4Dt}}-\sqrt{V_0t}\right) +\exp \left \lbrace y\sqrt{\frac{V_{0}}{D}}\right \rbrace\mathrm{erfc}\left( \frac{y}{\sqrt{4Dt}}+\sqrt{V_0t}\right)\right). \nonumber
\eeq
For $t$ small, using the expansion for the complementary error function for large arguments we compute the
numerator of the EMFPT (relation \ref{numerator}) as
\beq
s(t) &\sim& \left( 1-\frac{e^{-\frac{y^2}{4Dt}} \sqrt{4Dt}}{y \sqrt{\pi}}\frac{\left(e^{ -y\sqrt{\frac{V_{0}}{D}}}+e^{ y\sqrt{\frac{V_{0}}{D}}}\right)}{2}  \right)^n-\left(1-e^{ -y\sqrt{\frac{V_{0}}{D}}} \right)^n \nonumber\\
&\sim& 1 -\left(1-e^{ -y\sqrt{\frac{V_{0}}{D}}} \right)^n +\sum_{k=1}^n \binom{n}{k}\left(\frac{e^{-\frac{y^2}{4Dt}} \sqrt{4Dt}}{y \sqrt{\pi}}\frac{\left(e^{ -y\sqrt{\frac{V_{0}}{D}}}+e^{ y\sqrt{\frac{V_{0}}{D}}}\right)}{2}\right)^k. \nonumber
\eeq
This, leads to the following integral dominated for $t$ small when $n$ large,
\beq \label{formula2}
\bar\tau_{EMFPT}(n)&\sim& \int\limits_0^{\delta} \left[1-n \frac{\sqrt{4Dt} \exp \left \lbrace -\frac{y^2}{4Dt} \right\rbrace\left(e^{ -y\sqrt{\frac{V_{0}}{D}}}+e^{ y\sqrt{\frac{V_{0}}{D}}}\right)}{2y \sqrt{\pi}\left(1 -\left(1-e^{ -y\sqrt{\frac{V_{0}}{D}}} \right)^n\right)}\right]\, dt  \\
&\sim& \int\limits_0^{\infty} \exp \left\lbrace -n \frac{\sqrt{4Dt} \exp \left \lbrace -\frac{y^2}{4Dt} \right\rbrace \left(e^{ -y\sqrt{\frac{V_{0}}{D}}}+e^{ y\sqrt{\frac{V_{0}}{D}}}\right)}{2y \sqrt{\pi}\left(1 -\left(1-e^{ -y\sqrt{\frac{V_{0}}{D}}} \right)^n\right)} \right\rbrace \, dt,\nonumber
\eeq
and proceeding as in \cite{basnayake2019asymptotic}, we get
\beq \label{formula3}
\bar\tau_{EMFPT}(n)&\sim& \frac{y^2}{4D \log \displaystyle \left( \frac{n\left(e^{ -y\sqrt{\frac{V_{0}}{D}}}+e^{ y\sqrt{\frac{V_{0}}{D}}}\right)}{2\sqrt{\pi}\left(1 -\left(1-e^{ -y\sqrt{\frac{V_{0}}{D}}} \right)^n\right)}\right)}.
\eeq
Note that, when $V_0 = 0$, we recover the asymptotic formula for the case without killing and a Dirac-delta function as initial condition.
\subsection{Killing in a finite interval in half a line with initial point outside the interval}
We consider the diffusion of a particle that starts at a point $y$ outside the interval $[0,L]$. The pdf of that particle's trajectory
satisfies the equation
\beq\label{killing in an interval}
\frac{\partial p(x,t\,|\,y)}{\partial t}&=&D\frac{\partial^{2}
p(x,t\,|\,y)}{\partial x^{2}} - V \chi_{[0,L]}(x)  p(x,t\,|\,y)\quad\hbox{ on } \mathbf{R}_+ \\
p(x,0\,|\,y)&=&\delta(x-y) \nonumber \\
p(0,t\,|\,y)&=&0. \nonumber
\eeq
To compute the explicit solution, $p(x,t\,|\,\y)$, we Laplace
transform the equation with respect to $t$ and we obtain the equation
\beq
&&\frac{\partial ^2 u}{\partial x^2}(x,q) - \left( \frac{q+V}{D}\right) u(x,q) = 0 \,\,\,\,\,\text{for $x \in [0,L]$} \nonumber \\
&& \frac{\partial ^2 u}{\partial x^2}(x,q) - \left( \frac{q}{D}\right) u(x,q) = -\frac{1}{D}\delta (x-y) \,\,\,\,\,\text{for $x \in (L,+\infty)$}, \nonumber
\eeq
where $u(x,q) = \mathcal{L} \left( p(x,t\,|\,\y)\right)$, and the bounded solutions in $\mathbf{R}_+$ are in the form
\beq
&&u(x,q) = A \exp \left \lbrace -\sqrt{\frac{q+V}{D}}x \right \rbrace -A \exp \left \lbrace \sqrt{\frac{q+V}{D}}x \right \rbrace \,\,\,\,\,\text{for $x \in [0,L]$} \nonumber \\
&&u(x,q) = \frac{1}{\sqrt{4Dq}}  \exp \left \lbrace -\sqrt{\frac{q}{D}}|x-y| \right \rbrace+ B\exp \left \lbrace -\sqrt{\frac{q}{D}}|x+y| \right \rbrace\,\,\,\,\,\text{for $x \in (L,+\infty)$}. \nonumber
\eeq
We are looking for the solutions that are continuous at $x=L$ and its first derivative is also continuous at $x=L$, then solving the corresponding system we get
\beq
A &=& -\frac{e^{\sqrt{\frac{q}{D}}(L-y)}}{D\left( \left(\sqrt{\frac{q+V}{D}}-\sqrt{\frac{q}{D}}\right) e^{-\sqrt{\frac{q+V}{D}}L} +\left(\sqrt{\frac{q+V}{D}}+\sqrt{\frac{q}{D}}\right) e^{ \sqrt{\frac{q+V}{D}}L } \right)}, \nonumber \\
B &=& \displaystyle\frac{ \left(\sqrt{\frac{q+V}{D}}-\sqrt{\frac{q}{D}}\right) e^{ -\left(\sqrt{\frac{q+V}{D}}-2\sqrt{\frac{q}{D}}\right)L} -\left(\sqrt{\frac{q+V}{D}}-\sqrt{\frac{q}{D}}\right) e^{ \left(\sqrt{\frac{q+V}{D}}+2\sqrt{\frac{q}{D}}\right)L} }{\sqrt{4Dq}\left( \left(\sqrt{\frac{q+V}{D}}-\sqrt{\frac{q}{D}}\right)e^{ -\sqrt{\frac{q+V}{D}}L} +\left(\sqrt{\frac{q+V}{D}}+\sqrt{\frac{q}{D}}\right) e^{ \sqrt{\frac{q+V}{D}}L } \right)}. \nonumber
\eeq
Using relation (\ref{escapetau}), we have
\beq
\int_0^\infty J(t)dt=D\int_{0}^{t} \frac{\p p}{\p x}(x=0,t|\,y)\,dt = D \frac{\p u}{\p x}(0,0) =\frac{1}{\cosh\left( \sqrt{\frac{V}{D}}L\right)}. \nonumber
\eeq
For $t$ small, we have
\beq
\int_0^t J(s)ds &=& D\int_{0}^{t } \frac{\p p}{\p x}(x=0,s|\,y)\,ds  \nonumber \sim \int_{0}^{t} \left[ \mathcal{L}^{-1}_s\left( e^{-y\sqrt{\frac{q}{D}}}\right) -VL \mathcal{L}^{-1}_s\left(\frac{ e^{-y\sqrt{\frac{q}{D}}}}{\sqrt{4Dq}}\right) \right]ds \nonumber \\
&\sim& \mathrm{erfc}\left(\frac{y}{\sqrt{4Dt}}\right).\nonumber
\eeq
Then, we have
\beq
P_{\infty}=1-\left(1-\int_{0}^{\infty } J(t\,|\,\y)dt\right)^n = 1-\left(1-\frac{1}{\cosh\left( \sqrt{\frac{V}{D}}L\right)} \right)^n,\nonumber
\eeq
and
\beq
s(t) &\sim& \left( 1-\frac{e^{-\frac{y^2}{4Ds}} \sqrt{4Dt}}{y \sqrt{\pi}}\right)^n-\left(1\frac{1}{\cosh\left( \sqrt{\frac{V}{D}}L\right)} \right)^n \nonumber\\
&\sim& 1 -\left(1-\frac{1}{\cosh\left( \sqrt{\frac{V}{D}}L\right)} \right)^n +\sum_{k=1}^n \binom{n}{k}\left(\frac{e^{-\frac{y^2}{4Ds}} \sqrt{4Dt}}{y \sqrt{\pi}}\right)^k. \nonumber
\eeq
This, leads to the following integral dominated for $t$ small when $n$ large
\beq
\bar\tau_{EMFPT}(n)&\sim& \int\limits_0^{\delta} \left[1-n \frac{\sqrt{4Dt} \exp \left \lbrace -\frac{y^2}{4Dt} \right\rbrace}{y \sqrt{\pi}\left(1 -\left(1-\frac{1}{\cosh\left( \sqrt{\frac{V}{D}}L\right)} \right)^n\right)}\right]\, dt \nonumber \\
&\sim& \int\limits_0^{\infty} \exp \left\lbrace -n \frac{\sqrt{4Dt} \exp \left \lbrace -\frac{y^2}{4Dt} \right\rbrace }{y \sqrt{\pi}\left(1 -\left(1-\frac{1}{\cosh\left( \sqrt{\frac{V}{D}}L\right)} \right)^n\right)} \right\rbrace \, dt,\nonumber
\eeq
and proceeding as in \cite{basnayake2019asymptotic}, we get
\beq \label{formula4}
\displaystyle \bar\tau_{EMFPT}(n)&\sim& \frac{y^2}{4D \log \left( \frac{n}{\sqrt{\pi}\left(1 -\left(1-\frac{1}{ \cosh\left( \sqrt{\frac{V}{D}}L\right)} \right)^n\right)}\right)}.
\eeq
Note that, when $V= 0$, we recover the asymptotic formula for the case without a killing term and a Dirac-delta function as initial condition.
\subsection{Killing in a finite interval in half a line with initial point inside the interval}
In this case, we consider the diffusion of a particle that starts at a point $y$ inside the interval $[0,L]$, then the pdf of the particle's trajectory satisfies the equation (\ref{killing in an interval}) but when we apply the Laplace transform to this equation, we get
\beq
&&\frac{\partial ^2 u}{\partial x^2}(x,q) - \left( \frac{q+V}{D}\right) u(x) =  -\frac{1}{D}\delta (x-y) \hspace{1.1cm}\text{for $x \in [0,L]$} \nonumber \\
&&\frac{\partial ^2 u}{\partial x^2}(x,q) - \left( \frac{q}{D}\right) u(x) = 0 \hspace{3.5cm}\text{for $x \in (L,+\infty)$}, \nonumber
\eeq
where $u(x,q) = \mathcal{L} \left( p(x,t\,|\,\y)\right)$. Here, the bounded solutions in $\mathbf{R}_+$ are in the form
\beq
&&u(x,q) = A \left(\exp \left \lbrace -\sqrt{\frac{q+V}{D}}|x-y| \right \rbrace-\exp \left \lbrace -\sqrt{\frac{q+V}{D}}|x+y| \right \rbrace\right) \nonumber \\
&+&\left(A -\frac{1}{\sqrt{4D(q+V)}}\right) \left(\exp \left \lbrace \sqrt{\frac{q+V}{D}}|x-y| \right \rbrace-\exp \left \lbrace \sqrt{\frac{q+V}{D}}|x+y| \right \rbrace \right) \,\,\,\text{for $x \in [0,L]$} \nonumber \\
&&u(x,q) =  B\exp \left \lbrace -\sqrt{\frac{q}{D}}x \right \rbrace \,\,\,\text{for $x \in (L,+\infty)$}. \nonumber
\eeq
Because we are looking for the continuous solutions at $x=L$ with first derivative continuous at $x=L$, we can solve the corresponding system and we get
\beq
A &=&\scriptstyle \frac{- \left(\sqrt{\frac{q+V}{D}}+\sqrt{\frac{q}{D}}\right) \left(e^{\sqrt{\frac{q+V}{D}}(L-y)} -e^{\sqrt{\frac{q+V}{D}}(L+y) }\right) \frac{1}{\sqrt{4D(q+V)}}}{\left( \left(\sqrt{\frac{q+V}{D}}-\sqrt{\frac{q}{D}}\right)\left( e^{-\sqrt{\frac{q+V}{D}}(L-y)} -e^{ -\sqrt{\frac{q+V}{D}}(L+y)} \right) -\left(\sqrt{\frac{q+V}{D}}+\sqrt{\frac{q}{D}}\right) \left(e^{ \sqrt{\frac{q+V}{D}}(L-y)}-e^{\sqrt{\frac{q+V}{D}}(L+y)}\right) \right)}, \nonumber \\
B &=& \scriptstyle \frac{e^{-\sqrt{\frac{q}{D}}L}}{D\left( \left(\sqrt{\frac{q+V}{D}}-\sqrt{\frac{q}{D}}\right)\left( e^{-\sqrt{\frac{q+V}{D}}(L-y) } -e^{-\sqrt{\frac{q+V}{D}}(L+y) }\right) -\left(\sqrt{\frac{q+V}{D}}+\sqrt{\frac{q}{D}}\right) \left(e^{ \sqrt{\frac{q+V}{D}}(L-y)} -e^{ \sqrt{\frac{q+V}{D}}(L+y)} \right)\right)}. \nonumber
\eeq
Using relation (\ref{escapetau}), we have
\beq
\int_0^\infty J(t)dt=D\int_{0}^{t} \frac{\p p}{\p x}(x=0,t|\,y)\,dt = D \frac{\p u}{\p x}(0,0) = \exp \left \lbrace -y\sqrt{\frac{V}{D}}\right \rbrace. \nonumber
\eeq
For $t$ small, we get
\beq
\int_0^t J(s)ds &=& D\int_{0}^{t } \frac{\p p}{\p x}(x=0,s|\,y)\,ds  =\int_{0}^{t}  \mathcal{L}^{-1}_s\left( e^{-y\sqrt{\frac{q}{D}}}\right) ds =\mathrm{erfc}\left(\frac{y}{\sqrt{4Dt}}\right). \nonumber
\eeq
Then, as in the case for the uniform killing, we get the asymptotic formula
\beq \label{uniformkilling}
\displaystyle \bar\tau_{EMFPT}(n)&\sim& \frac{y^2}{4D \log \left( \displaystyle \frac{n\left(e^{-y\sqrt{\frac{V}{D}}}+e^{ y\sqrt{\frac{V}{D}}}\right)}{2\sqrt{\pi}\left(1 -\left(1-e^{ -y\sqrt{\frac{V}{D}}} \right)^n\right)}\right)}.
\eeq
\newpage
\normalem
\bibliographystyle{plain}
\bibliography{bib_for_killing}
\end{document}